\documentclass[letterpaper]{article} \usepackage{aaai25}
\usepackage{subcaption}
\usepackage{times}  \usepackage{helvet}  \usepackage{courier}  \usepackage[hyphens]{url}  \usepackage{graphicx}    \usepackage{natbib}  \usepackage{caption} \usepackage{soul}
\usepackage{todonotes}
\usepackage{enumitem}
\usepackage{tabularx}
\usepackage{booktabs}
\usepackage{adjustbox}
\usepackage{fontawesome5}
\usepackage[hang,flushmargin]{footmisc}
\newcommand{\multicomment}[1]{}

    \usepackage{algorithm}
\usepackage{algorithmic}

\usepackage{newfloat}
\usepackage{listings}
\DeclareCaptionStyle{ruled}{labelfont=normalfont,labelsep=colon,strut=off} 
\floatstyle{ruled}
\newfloat{listing}{tb}{lst}{}
\floatname{listing}{Listing}
\newcommand{\dbField}[1]{{\small\texttt{#1}}}

\title{A Year of the DSA Transparency Database: What it (Does Not) Reveal\\About Platform Moderation During the 2024 European Parliament Election}
\author {
Gautam Kishore Shahi\textsuperscript{\rm 1},
    Benedetta Tessa \textsuperscript{\rm 2,3},
    Amaury Trujillo \textsuperscript{\rm 2},
    Stefano Cresci \textsuperscript{\rm 2}
}
\affiliations {
\textsuperscript{\rm 1}University of Duisburg-Essen\\
    \textsuperscript{\rm 2}IIT-CNR\\
    \textsuperscript{\rm 3}University of Pisa\\
    gautam.shahi@uni-due.de, benedetta.tessa@iit.cnr.it, 
    benedetta.tessa@phd.unipi.it, 
    amaury.trujillo@iit.cnr.it,
    stefano.cresci@iit.cnr.it
}

\begin{document}

\maketitle

\begin{abstract}
Social media platforms face heightened risks during major political events; yet, how platforms adapt their moderation practices in response remains unclear. The Digital Services Act Transparency Database offers an unprecedented opportunity to systematically study content moderation at scale, enabling researchers and policymakers to assess platforms' compliance and effectiveness. Herein, we analyze 1.58 billion self-reported moderation actions taken by eight large social media platforms during an extended period of eight months surrounding the 2024 European Parliament elections. Our findings reveal a lack of adaptation in moderation strategies, as platforms did not exhibit significant changes in their enforcement behaviors surrounding the elections. This raises concerns about whether platforms adapted their moderation practices at all, or if structural limitations of the database concealed possible adjustments. Moreover, we found that noted transparency and accountability issues persist nearly a year after initial concerns were raised. These results highlight the limitations of current self-regulatory approaches and underscore the need for stronger enforcement and data access mechanisms to ensure that online platforms uphold their responsibility in safeguarding democratic processes.
\end{abstract}

\section{Introduction}
The European Parliament elections are paramount in shaping European Union (EU) politics. Held every five years, they allow citizens to elect their representatives for EU's legislative power, with the newly elected Parliament in turn electing the President of the European Commission. The most recent European Parliament elections took place in 2024---a year with an extraordinary number of national elections worldwide---from June 6th to 9th across all EU countries. The President of the commission was later elected via a secret ballot on July 18th.
Such national and supra-national elections have a major impact on social media platforms, which play a crucial role in hosting political campaigns and disseminating election-related information~\cite{rho2020political,papakyriakopoulos2023upvotes,shahi2024agenda}. For example, politicians often engage with citizens and encourage political learning as well as electoral participation through their social media channels~\cite{cresci2014criticism,fatema2022social,kim2022observation,bene2022keep}. At the same time however, online electoral discourse can also be targeted by information tampering actions orchestrated to gain pre-electoral consent~\cite{chen2022social}. These activities include information manipulation~\cite{cinelli2020limited,nizzoli2021coordinated,matatov2022stop,hristakieva2022spread}, the use of deepfakes~\cite{haq2024,diakopoulos2021anticipating}, targeted harassment~\cite{hua2020characterizing}, as well as opaque and unfair political advertising~\cite{bar2024systematic}.

These phenomena are examples of the risks associated with periods of increased online political activity, where critical interests---both political and economic---come into play.
In response, social media platforms attempt to mitigate the issues through content moderation~\cite{gillespie2018custodians}. During electoral periods, online platforms may intensify efforts to ensure trustworthy online discourse through the enforcement of multiple content and account moderation actions~\cite{pierri2023does,majo2021role}.

Content moderation has recently gained increased relevance not only for platform users but also for European regulators. In October 2022, the European Union enacted the Digital Services Act (DSA) to regulate online platforms and foster a more transparent, inclusive, and safe digital environment~\cite{eu2020DSA}. Among its requirements, the DSA obliges large online platforms to report all their moderation actions within the EU by submitting clear, detailed, and timely \textit{statement of reasons} (SoRs) to the DSA Transparency Database (\texttt{DSA-TDB})\footnote{\url{https://transparency.dsa.ec.europa.eu/}}---an open and centralized repository hosted by the European Commission~\cite{trujillo2023dsa,kaushal2024automated}. Operational since September 2023, the \texttt{DSA-TDB} represents an unprecedented tool for transparency and promised to revolutionize the observability of online platforms.
For this reason, a few early works have analyzed the initial information that platforms submitted to the database during its first months of operation~\cite{trujillo2023dsa,dergacheva2023one,aspromonte-etal-2024-llms,kaushal2024automated,drolsbach2024content}. However, these uncovered compliance deficiencies by platforms and highlighted significant issues in the structure of the database itself, all of which limit its overall usefulness and reliability~\cite{trujillo2023dsa,kaushal2024automated}. 

Here, we carry out the most extensive analysis of the \texttt{DSA-TDB} to date, almost one year after its initial release. We consider a broad observation period of ten months surrounding the 2024 EU elections and we explore the self-reported moderation actions of the eight largest social media platforms in the EU. By analyzing 1.58B SoRs  in a politically critical time, we seek to understand whether the \texttt{DSA-TDB} has lived up to its transparency promise, assessing if and how the reported moderation practices of large social media platforms changed in response to the heightened integrity risks. Additionally, our extensive exploratory analysis allows assessing whether the shortcomings reported in previous studies have been addressed. This work thus answers the following research questions:
\begin{itemize}
    \item \textbf{RQ1:} \textit{How did self-reported moderation practices change during the European electoral period?} We aim to identify possible notable shifts in content moderation practices before and after the 2024 EU elections, such as changes in the volume and type of moderated content. 
    \item \textbf{RQ2:} \textit{To what extent has the reliability and consistency of the database improved since its initial release?} Initial analyses identified significant issues that limited the practical utility of the database as a transparency tool. Our study revisits these concerns with a larger and recent dataset, assessing progress and persistent challenges in achieving meaningful platform transparency.
\end{itemize}

Our results contribute to advancing the understanding of transparency mechanisms in digital governance, inform future regulatory decisions, and provide a timely resource for policymakers, scholars, and platforms aiming to foster greater integrity and accountability in online spaces.
 \section{Related work}
The introduction of the European \texttt{DSA-TDB} has marked a significant milestone in digital legislation worldwide, aiming to enhance fairness and transparency in online governance. Hence, despite being launched only in September 2023, it has already been the focus of multiple studies assessing its effectiveness and the compliance of involved platforms. For instance, an initial study analyzed a 10-day sample covering all platforms to evaluate whether the database met its transparency and fairness objectives~\cite{kaushal2024automated}. While this study acknowledged the value of the insights into moderation decisions, it also identified inconsistencies and non-compliance issues arising from the self-reported nature of the data. Other analyses further highlighted heterogeneity in content moderation practices, including differences in the types of moderated content, the application of visibility restrictions, and the reported use of automation~\cite{drolsbach2024content,dergacheva2023one}. Extending these early investigations, a 100-day study found that while platforms formally comply with the requirements of the \texttt{DSA-TDB}, they often omit key optional details in their SoRs, limiting the database’s practical utility~\cite{trujillo2023dsa}. This study also revealed substantial discrepancies in self-reported moderation practices across platforms, suggesting varying levels of adherence to the intended structure of the \texttt{DSA-TDB}. Moreover, cross-checking the database against platforms' own transparency reports exposed significant inconsistencies in the submitted data. Beyond direct assessments of the database’s quality, other work has explored ways to enhance its usability. For instance,~\citet{aspromonte-etal-2024-llms} employed a multi-agent  system based on large language models (LLMs) to link SoRs to the corresponding sections of platforms’ Terms of Service. Their findings suggest that LLMs can provide valuable contextualization, improving user understanding of moderation decisions and potentially increasing engagement with the DSA.

While these studies offer critical insights, they were conducted within the first months after the \texttt{DSA-TDB}’s release. Now---more than a year later---it is crucial to revisit these findings, using a larger dataset to assess whether the database has improved in consistency, completeness, and overall transparency, or whether the initial concerns remain unresolved. These issues are particularly relevant during a major continental election, in which moderation practices may experience major changes.

\section{Data}
Our dataset consists of 1.58B SoRs that we collected from the publicly-available \texttt{DSA-TDB}.\footnote{\url{https://transparency.dsa.ec.europa.eu/explore-data/download}} The SoRs cover all self-reported moderation actions that eight very large social media platforms took in the EU between March and October 2024. The chosen time frame covers approximately 14 weeks before and 20 weeks after the election days, allowing a thorough analysis of the possible shift in moderation practices before and after the electoral period. In detail, our data includes 646.1M SoRs from TikTok, 300.8M from Instagram, 260.2M from Facebook, 81.9M from Pinterest, 36.3M from YouTube, 2.3M from Snapchat, 628K from X, and 293K from LinkedIn. As explained in the database's official documentation, each SoR is composed of multiple fields.\footnote{\url{https://transparency.dsa.ec.europa.eu/page/documentation}} All those utilized in this study are described in Appendix Table~\ref{tab:dsa-fields}.

 \begin{figure*}[t]
\centering
    \begin{subfigure}{0.25\textwidth}\includegraphics[width=\textwidth]{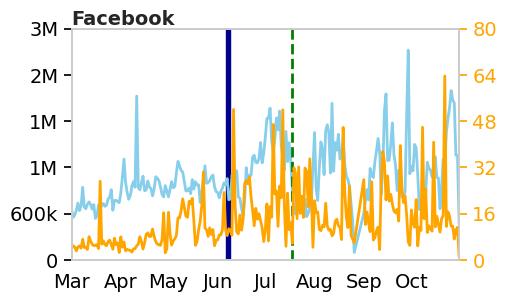}\label{fig:timeline_facebook}\end{subfigure}\begin{subfigure}{0.25\textwidth}\includegraphics[width=\textwidth]{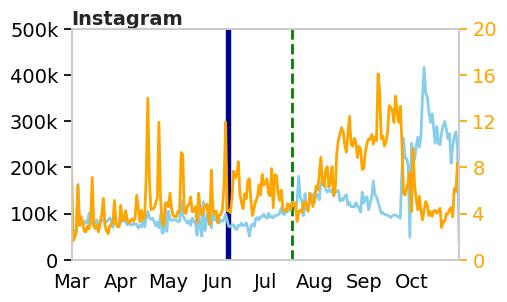}\label{fig:timeline_instagram}\end{subfigure}\begin{subfigure}{0.24\textwidth}\includegraphics[width=\textwidth]{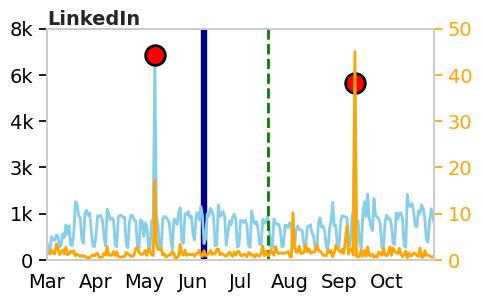}\label{fig:timeline_linkedin}\end{subfigure}\begin{subfigure}{0.26\textwidth}\includegraphics[width=\textwidth]{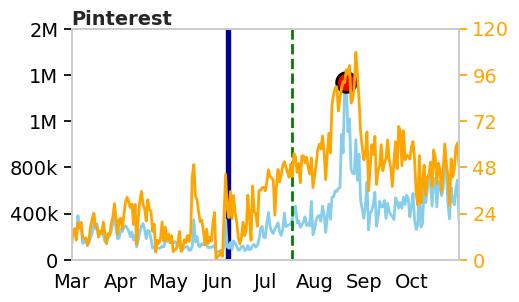}\label{fig:timeline_pinterest}\end{subfigure}\\
    \begin{subfigure}{0.25\textwidth}\includegraphics[width=\textwidth]{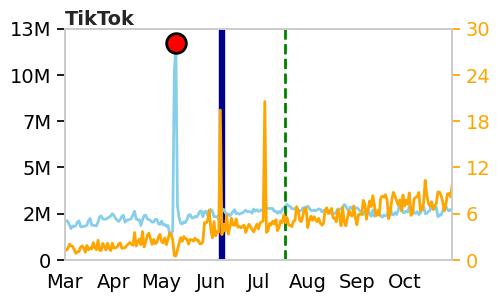}\label{fig:timeline_tiktok}\end{subfigure}\begin{subfigure}{0.24\textwidth}\includegraphics[width=\textwidth]{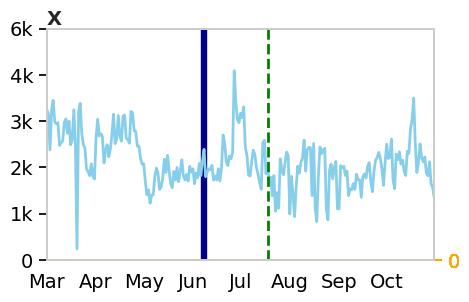}\label{fig:timeline_x}\end{subfigure}\begin{subfigure}{0.26\textwidth}\includegraphics[width=\textwidth]{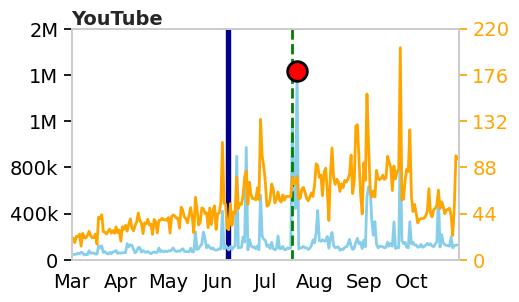}\label{fig:timeline_youtube}\end{subfigure}\begin{subfigure}{0.25\textwidth}\includegraphics[width=\textwidth]{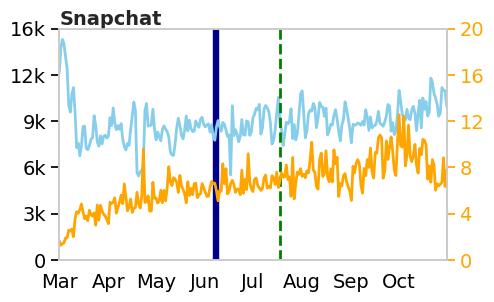}\label{fig:timeline_snapchat}\end{subfigure}\caption{Daily time series of the number of moderation actions (cyan-colored) and their average delay (orange-colored). The blue vertical band indicates the Parliament elections days (6th--9th June), while the green vertical dashed line indicates the President election day (18th July). The red circles highlight the subset of anomalies that we analyzed in detail.}
\label{fig:timeline}
\end{figure*}

\begin{figure*}[t]
\centering
    \begin{subfigure}{0.25\textwidth}\includegraphics[width=\textwidth]{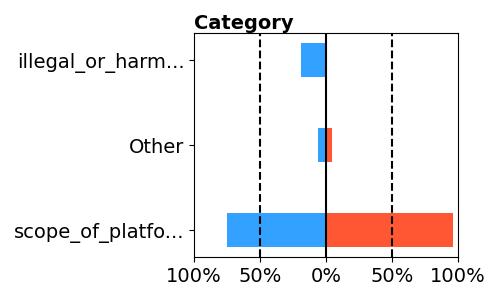}\end{subfigure}\begin{subfigure}{0.25\textwidth}\includegraphics[width=\textwidth]{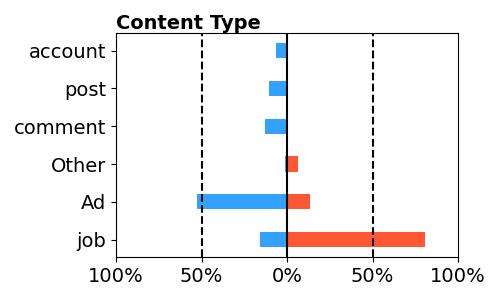}\end{subfigure}\begin{subfigure}{0.25\textwidth}\includegraphics[width=\textwidth]{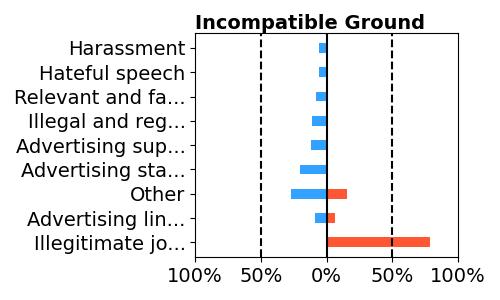}\end{subfigure}\begin{subfigure}{0.25\textwidth}\includegraphics[width=\textwidth]{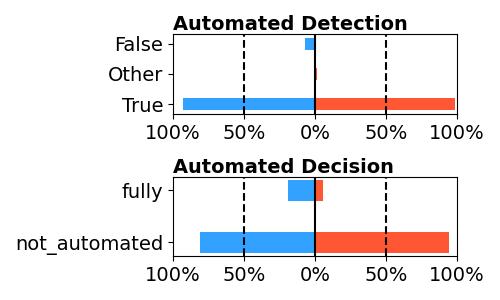}\end{subfigure}\caption{Comparison between the SoRs (right-aligned, red-colored) that caused the moderation \textit{volume} anomaly reported by \faLinkedin~\textbf{LinkedIn} on \textbf{May 8, 2024} and the SoRs (left-aligned, blue-colored) by the same platform from the surrounding routine days.}
    \label{anomaly-linkedin-volume}
\end{figure*}

\section{Analyses and Results}
\subsection{Trends in moderation actions}
To address RQ1, we first examine whether platforms adjusted the volume or timeliness of their moderation actions in response to the heightened integrity risks of the 2024 EU elections. A surge in moderation activity or a reduction in moderation delays could indicate increased vigilance, whereas stationarity in these metrics might suggest that moderation remained unchanged despite the elections. To analyze these trends, we constructed daily time series of moderation decisions made by each platform. Additionally, we computed daily time series of SoR moderation delays, defined as the number of days between the content’s creation of the SoR and its eventual moderation action. For each platform, moderation delays were averaged on a daily basis. Figure~\ref{fig:timeline} presents both the volume (cyan-colored) and delay (orange-colored) time series, allowing us to assess whether notable shifts in moderation practices occurred during the electoral period compared to routine periods.

Figure~\ref{fig:timeline} reveals substantial heterogeneity across platforms in both moderation volume and delay patterns. Some platforms take millions of moderation actions daily, while others only a few thousands. Certain platforms, like X, show stable trends in moderation delay, while others, like LinkedIn, have regular fluctuations in moderation volume. Moreover, some platforms present frequent and sharp variations, indicating dynamic or event-driven moderation policies. The differences persist even when accounting for each platform’s number of active users~\cite{trujillo2023dsa}. However, no clear or systematic shift in moderation behavior is observed during the electoral period, indicated by the vertical blue and green lines corresponding to the European Parliament and Presidential elections, respectively. This suggests that, at a broad level, moderation activity remained relatively consistent before, during, and after the elections. Nonetheless, nearly all platforms exhibit distinct spikes in either the volume or delay of moderation at various points in time. To determine if these anomalies are linked to election-related tampering, we conduct a focused analysis on a subset of these peaks. The anomalies selected for further scrutiny are marked with red dots in Figure~\ref{fig:timeline}.

\begin{figure*}[t]
\centering
    \begin{subfigure}{0.25\textwidth}\includegraphics[width=\textwidth]{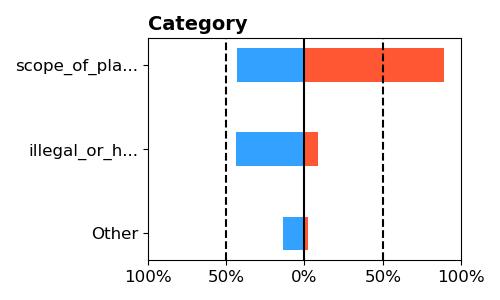}\end{subfigure}\begin{subfigure}{0.25\textwidth}\includegraphics[width=\textwidth]{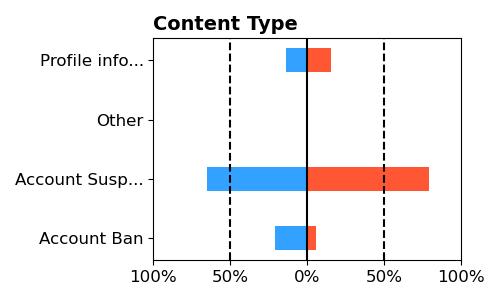}\end{subfigure}\begin{subfigure}{0.25\textwidth}\includegraphics[width=\textwidth]{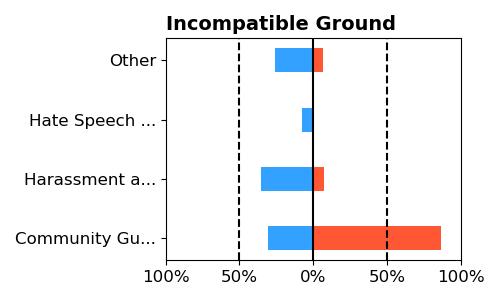}\end{subfigure}\begin{subfigure}{0.25\textwidth}\includegraphics[width=\textwidth]{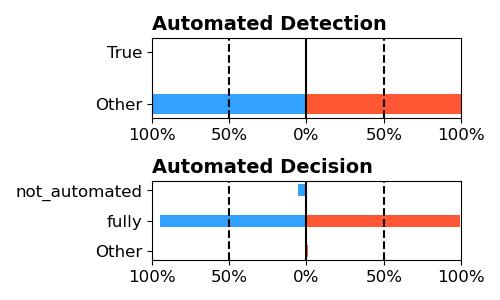}\end{subfigure}\caption{Comparison between the SoRs (right-aligned, red-colored) that caused the moderation \textit{volume} anomaly reported by \faTiktok~\textbf{TikTok} on \textbf{May 10, 2024} and the SoRs (left-aligned, blue-colored) by the same platform from the surrounding routine days.}
    \label{anomaly-tiktok-volume}
\end{figure*}

\begin{figure*}[t]
\centering
   \begin{subfigure}{0.25\textwidth}\includegraphics[width=\textwidth]{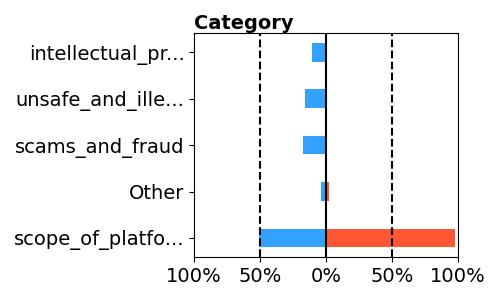}\end{subfigure}\begin{subfigure}{0.25\textwidth}\includegraphics[width=\textwidth]{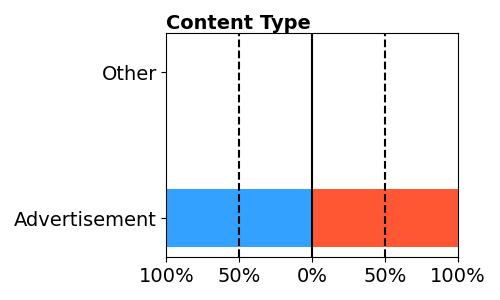}\end{subfigure}\begin{subfigure}{0.25\textwidth}\includegraphics[width=\textwidth]{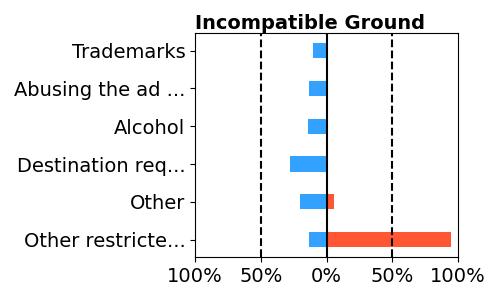}\end{subfigure}\begin{subfigure}{0.25\textwidth}\includegraphics[width=\textwidth]{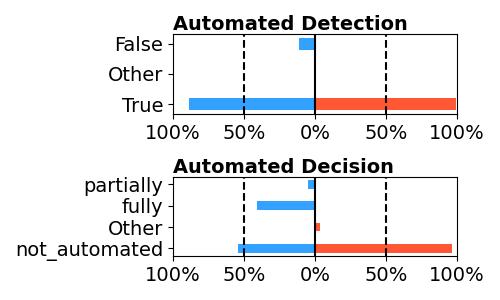}\end{subfigure}\caption{Comparison between the SoRs (right-aligned, red-colored) that caused the moderation \textit{volume} anomaly reported by \faYoutube~\textbf{YouTube} on \textbf{July 21, 2024} and the SoRs (left-aligned, blue-colored) by the same platform from the surrounding routine days.}
    \label{anomaly-youtube-volume}
\end{figure*}

\subsection{Anomalies in moderation actions}
The time series analysis revealed no clear shifts in moderation trends. However, the presence of sharp spikes in moderation volume or delay might indicate that changes have occurred in a more localized manner. For example, these anomalies could indicate mass election-related moderation events, such as coordinated enforcement actions or ban waves, that are not reflected in the broader trends~\cite{decook2022r}. 
To investigate the nature of the selected anomalies, we compared the characteristics of the SoRs within each spike to those issued by the same platform in the days before and after the spike. This comparative analysis aims to identify possible marked shifts in the attributes of the SoRs that could explain the underlying moderation decisions. To this end, we are particularly interested in the use of specific SoR attributes and values designed to indicate election-related tampering (e.g., the predefined category \dbField{negative\_effects on\_civic\_discourse\_or\_elections}). Among the available ones, we analyze the type of infringement (\dbField{category}), the type of moderated content (\dbField{content\_type}), the specific reason for incompatibility with platform policies (\dbField{incompatible\_ground}), and the use of automation in moderation---distinguishing between \dbField{automated\_detection} and \dbField{automated\_decision}. For clarity and brevity, in the following analyses we show a subset of all attribute values---those mostly used by each platform and those for which we measured noticeable differences. Appendix Section ``Attributes values'' provides more information on attribute value selection. For each one, we visualize the differences using diverging bar charts, where left-aligned blue bars represent SoRs from routine (non-peak) days, and right-aligned red bars correspond to SoRs associated with the anomalies.

\noindent\textbf{\faLinkedin~LinkedIn---May 8, 2024.} Figure~\ref{anomaly-linkedin-volume} reveals that the SoRs associated with this anomaly differ significantly from routine moderation cases, particularly in the type of moderated content and the grounds for infringement. While LinkedIn mostly moderates advertisements, on May 8, 2024, the platform primarily targeted illegitimate job offers, suggesting a distinct shift in enforcement focus on that day. However, the available data do not provide further details on the rationale behind the moderation of such job offers, nor do they offer clear indications that would allow referring this spike to the electoral context.

\noindent\textbf{\faTiktok~TikTok---May 10, 2024.} Figure~\ref{anomaly-tiktok-volume} indicates that this anomaly stems from the mass suspension of accounts deemed to be operating outside of TikTok’s intended scope and in violation of its community guidelines. However, the information provided in these SoRs is highly generic, offering little insight into the specific reasons behind this moderation surge or whether it is directly linked to the electoral context. 

\noindent\textbf{\faYoutube~YouTube---July 21, 2024.} Figure~\ref{anomaly-youtube-volume} shows that this moderation spike primarily targeted advertisements flagged for being outside YouTube’s scope and, specifically, related to ``other restricted businesses.'' Furthermore, unlike YouTube’s usual moderation processes, all decisions in this case were issued automatically. Interestingly, some type of restricted business could be election-related,\footnote{\url{https://support.google.com/adspolicy/answer/6368711?hl=en}} such as businesses related to ``Government documents and official services.'' However, the SoRs submitted by YouTube do not specify which type of restricted businesses they are related to. As such, akin to the previous anomalies, the limited information provided makes it difficult to determine whether the moderation actions were directly related to the election.

\begin{figure*}[t]
\centering
    \begin{subfigure}{0.25\textwidth}\includegraphics[width=\textwidth]{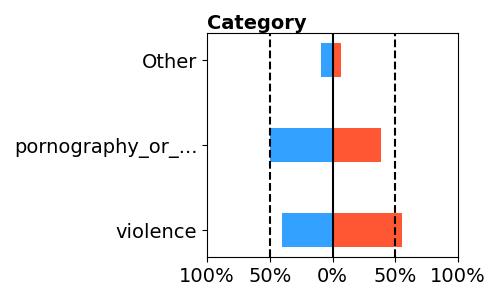}\end{subfigure}\begin{subfigure}{0.25\textwidth}\includegraphics[width=\textwidth]{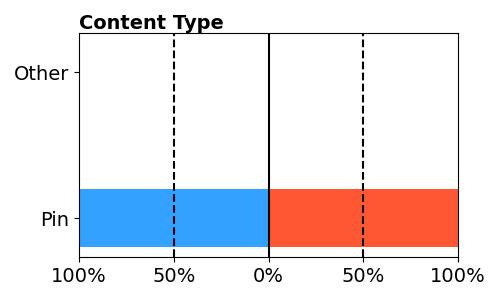}\end{subfigure}\begin{subfigure}{0.25\textwidth}\includegraphics[width=\textwidth]{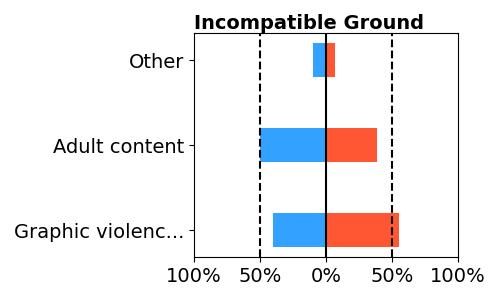}\end{subfigure}\begin{subfigure}{0.25\textwidth}\includegraphics[width=\textwidth]{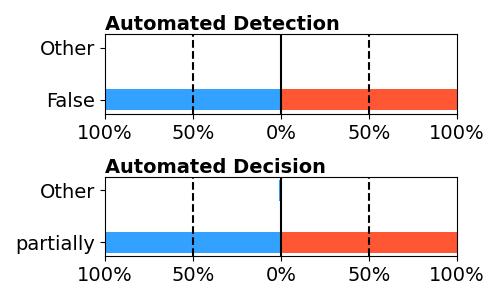}\end{subfigure}\caption{Comparison between the SoRs (right-aligned, red-colored) that caused the moderation \textit{volume} anomaly reported by \faPinterest~\textbf{Pinterest} on \textbf{August 21, 2024} and the SoRs (left-aligned, blue-colored) by the same platform from the surrounding routine days.}
    \label{fig:anomaly-pinterest-volume}
\end{figure*}

\begin{figure*}[t]
\centering
    \begin{subfigure}{0.25\textwidth}\includegraphics[width=\textwidth]{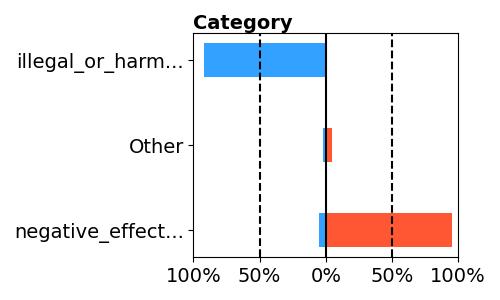}\end{subfigure}\begin{subfigure}{0.25\textwidth}\includegraphics[width=\textwidth]{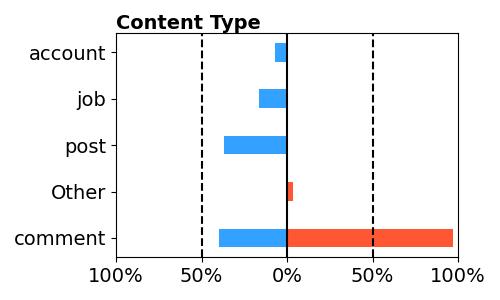}\end{subfigure}\begin{subfigure}{0.25\textwidth}\includegraphics[width=\textwidth]{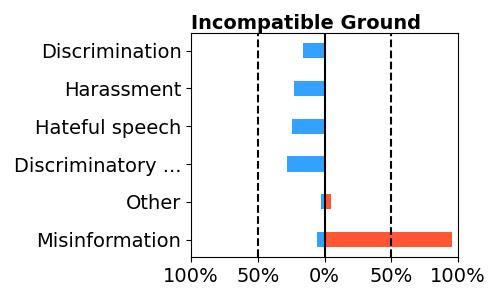}\end{subfigure}\begin{subfigure}{0.25\textwidth}\includegraphics[width=\textwidth]{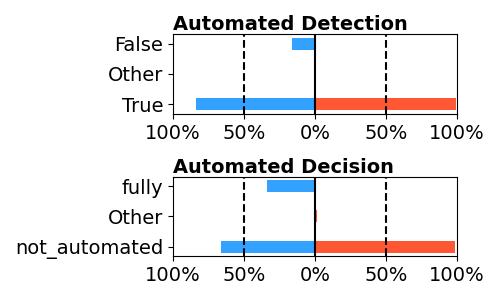}\end{subfigure}\caption{Comparison between the SoRs (right-aligned, red-colored) that caused the moderation \textit{delay} anomaly reported by \faLinkedin~\textbf{LinkedIn} on \textbf{September 11, 2024} and the SoRs (left-aligned, blue-colored) by the same platform from the surrounding routine days.}
    \label{fig:anomaly-linkedin-delay}
\end{figure*}

\noindent\textbf{\faPinterest~Pinterest---August 21, 2024.} Figure~\ref{fig:anomaly-pinterest-volume} surfaces minimal differences between the SoRs related to the moderation spike and the regular ones. The few differences involve a larger fraction of moderation actions targeted at graphic violence rather than at pornographic material. The information available from these SoRs suggests that this moderation spike was not related to the election.

\noindent\textbf{\faLinkedin~LinkedIn---September 11, 2024.} Figure~\ref{fig:anomaly-linkedin-delay} presents the most compelling anomaly in our analysis, revealing a moderation spike driven by a surge in comments flagged for election-related misinformation. A key indicator of its relevance is the explicit use of the \dbField{negative\_effects on\_civic\_discourse\_or\_elections} field---an unusually specific designation compared to the more generic moderation categories typically employed. Additionally, this anomaly stands out due to its lower reliance on automation, suggesting a more deliberate review process. We also note that this spike appears in the moderation delay time series, indicating that the moderated content was not recent but had been posted weeks earlier. Although the moderation actions occurred in September---after the electoral period---the delay of approximately 45 days traces the original publication of the moderated comments back to mid July. This timing is highly significant, as it falls between the European Parliament and Presidential elections, a period of heightened political discourse and potential misinformation risks. The distinct characteristics of this anomaly, both in terms of timing and the specificity of the moderation labels used, strongly suggest that this spike was indeed election-related. This case highlights the importance of granular labeling practices within the \texttt{DSA-TDB}, as LinkedIn’s use of precise categories rather than broad classifications (e.g., \dbField{scope\_of\_platform\_service}), enabled us to draw this conclusion with confidence.

Overall, our analysis highlights a significant limitation in assessing the observed moderation anomalies in relation to the European elections. In most cases, we were unable to confidently determine whether these surges in moderation actions were linked to the electoral context or driven by other factors. This uncertainty primarily stems from the lack of detailed information in the SoRs, which manifests in two key ways: \textit{(i)} the frequent use of generic values in mandatory fields, offering little insight into the specific rationale behind moderation decisions, and \textit{(ii)} the underutilization of optional fields, which could have provided crucial additional context.

\begin{figure*}[t]
    \centering
    \includegraphics[width=\textwidth]{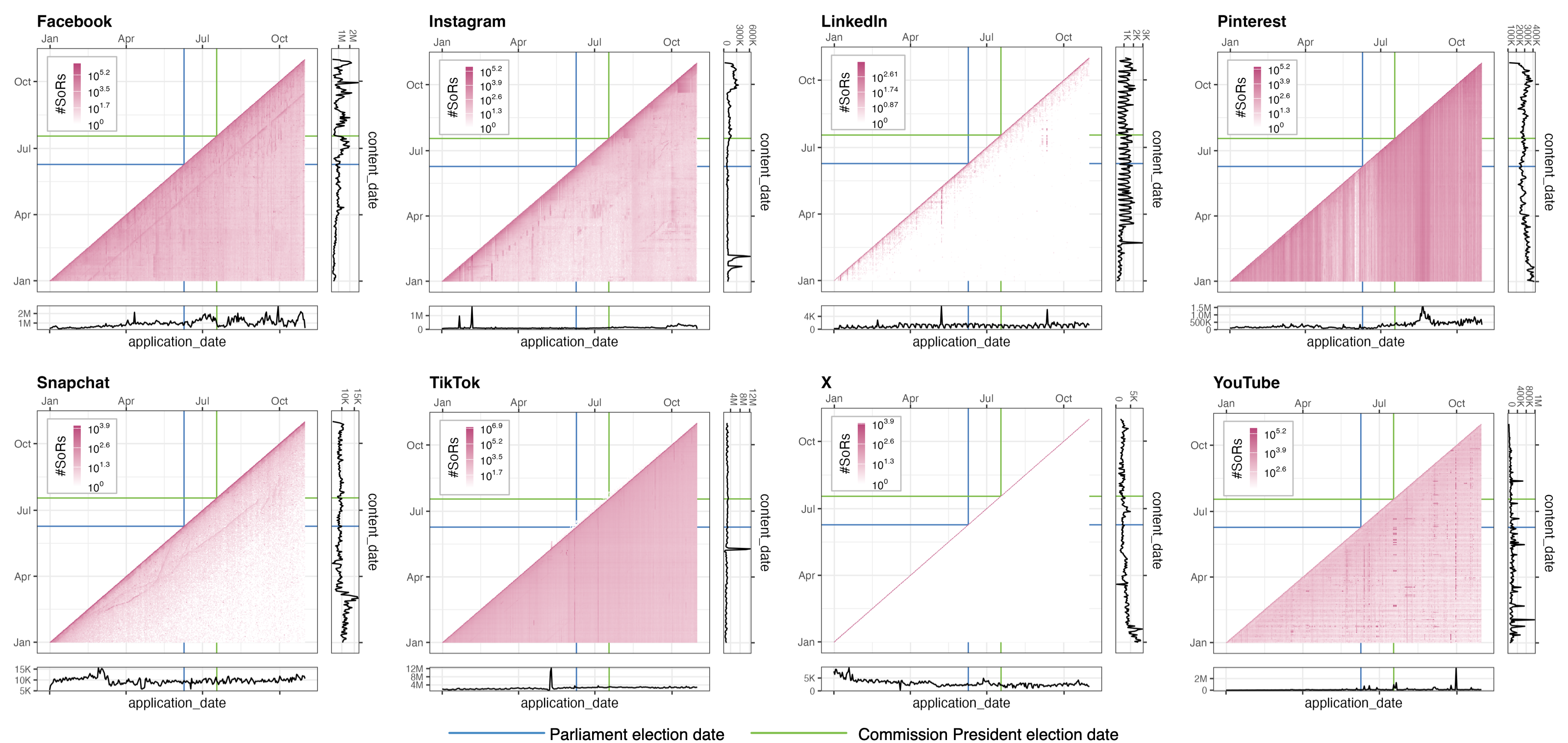}
    \caption{Analysis of moderation delays. For each platform, the heatmaps show the relationship, on a logarithmic scale, between the date when content was moderated (\textit{x} axis) and the date when the same content was published (\textit{y} axis). Blue lines indicate the Parliament elections days, while green lines indicate the President election day.}
\label{fig:delay_analysis}
\end{figure*}

\subsection{Delays in moderation actions}
The analysis of moderation trends revealed no significant changes in moderation volume during the electoral period. However, the orange-colored moderation delay time series shown in Figure~\ref{fig:timeline} reveal a steady increase in average daily delay for all platforms, except for X and LinkedIn. The increase is particularly evident in the post-electoral period, starting in late August and early September 2024. This pattern suggests that while moderation activity remained steady during the elections, platforms may have begun retrospectively moderating election-period content only after the elections had concluded. This delayed response could explain both the absence of noticeable shifts during the electoral period and the rise in moderation delays afterward. To investigate this hypothesis, we conducted a deeper analysis of moderation delays, examining whether content posted during the elections was moderated at a later stage.

Figure~\ref{fig:delay_analysis} presents the joint and marginal distributions of moderation date (\textit{x} axis) and content publication date (\textit{y} axis) for each platform. The central heatmaps show the relationship between these dates, on a logarithmic scale. Points along the main diagonal indicate no moderation delay (i.e., content moderated on the same day it was published), while points further below the diagonal represent moderation actions targeting older content. In this visualization, no point should ever lay above the main diagonal, as that would correspond to content moderated before being published. Figure~\ref{fig:delay_analysis} reveals highly heterogeneous moderation behaviors across platforms. LinkedIn and X consistently moderate content close to its publication date, with X exhibiting an extreme case, reporting zero moderation delay in all its submitted SoRs. YouTube, Instagram, and Snapchat primarily focus on recent content but occasionally moderate older posts. In contrast, Facebook and TikTok display a more uniform distribution, with moderation delays more evenly spread across time. Pinterest stands out, showing a moderation pattern largely independent of publication date, as indicated by the vertical lines in its heatmap. Other notable patterns are the dark-colored diagonal lines visible in Facebook's and Snapchat's heatmaps. Facebook, in particular, shows a consistent one-month lag for certain moderation actions, possibly reflecting batch reviews or scheduled automated moderation processes. Similar, though less pronounced, patterns appear in Instagram and YouTube, hinting at structured moderation processes.

Beyond these observations, Figure~\ref{fig:delay_analysis} allows us to test our initial hypothesis. If platforms had disproportionately moderated content after the electoral period, we would expect to see a clear pattern in the heatmaps---namely, a concentration of moderation actions targeting content published around the Parliament and President elections, but occurring predominantly from early September onward. However, no such pattern emerges in the data. The only notable exception is LinkedIn, where moderation of election-related misinformation comments---as discussed in the previous section---left a visible mark in the platform’s heatmap. Apart from this case, we find no evidence that moderation actions systematically targeted electoral-period content after the elections had concluded. In summary, with the exception of LinkedIn, our analyses of moderation trends, anomalies, and delays do not indicate any significant shifts in moderation practices before, during, or after the electoral period across the analyzed platforms.

\begin{figure}[t]
    \centering
    \includegraphics[width=0.6\columnwidth]{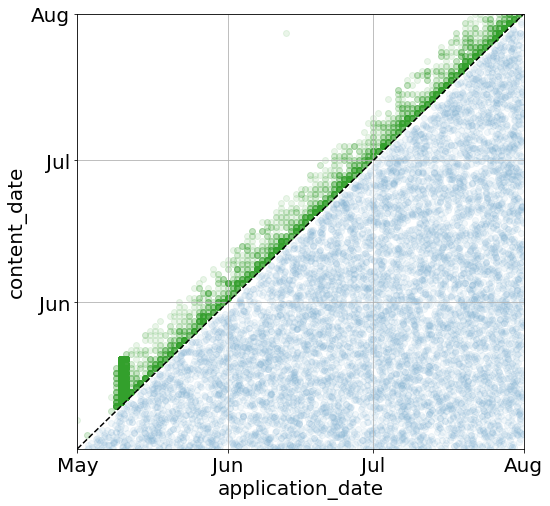}
    \caption{Highlight of 12,010 SoRs from TikTok that erroneously report content being moderated before it was published. The erroneous SoRs are green-colored while the correct ones are blue-colored.}
    \label{fig:tiktok_err}
\end{figure}

\subsection{Quality of platform-shared data}
In the months following the launch of the \texttt{DSA-TDB}, multiple studies uncovered important issues with the quality and consistency of the data submitted by platforms~\cite{trujillo2023dsa,kaushal2024automated,drolsbach2024content,papaevangelou2024content}. Given that the database was designed to promote transparency and accountability, these early shortcomings raised concerns about its reliability. Now---one year after its implementation---we revisit these findings by analyzing our recent dataset to answer RQ2 and determine whether the situation has improved. If the same inconsistencies persist, this would call into question not only the accuracy of the reported data but also the broader effectiveness of the database as a tool for regulatory oversight.

\begin{figure*}[t]
    \centering
    \includegraphics[width=0.8\textwidth]{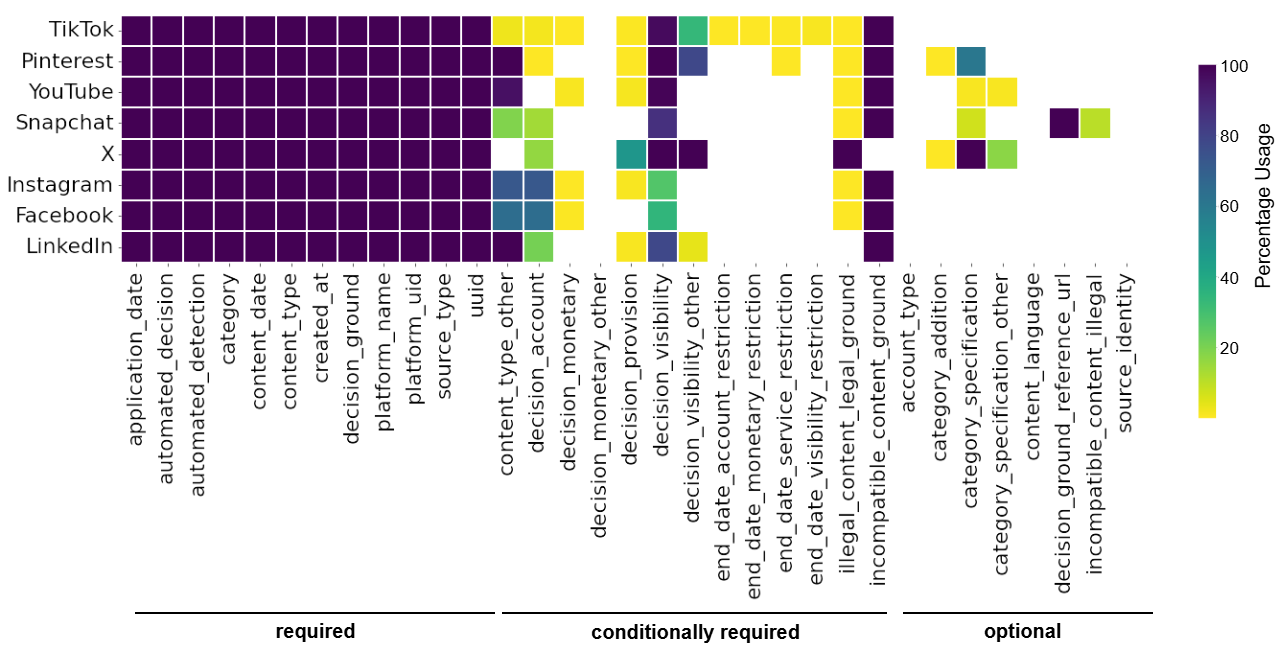}
    \caption{Heatmap showing the frequency with which each database attribute is used by each platform. Attributes are grouped based on whether they are required, conditionally required, or optional. Platforms are displayed in descending order of attribute utilization.}
    \label{fig:field_usage}
\end{figure*}

\noindent\textbf{Erroneous database records.} To assess the quality and reliability of the data, we first searched for obvious errors. Our analysis uncovered two issues: \textit{(i)} duplicate records, and \textit{(ii)} records indicating that content was supposedly moderated before it was even published. The first issue affected a small minority of records submitted by Facebook and Pinterest, and involves SoRs with identical Universally Unique Identifiers (UUIDs), which should be unique by design. In every case we identified, these duplicate SoRs were completely identical across all fields. The second issue is more concerning: several SoRs by TikTok, Facebook, Snapchat, and LinkedIn reported moderation actions occurred before the corresponding content was published. Figure~\ref{fig:tiktok_err} highlights this pattern in a subset of TikTok data from May to August 2024, where green-colored records erroneously indicate preemptive moderation. Although these errors represent a small fraction of the overall SoRs---for example, TikTok submitted over 12k erroneous SoRs on May 10, 2024, accounting for $\sim$0.1\% of its total that day---care is needed when analyzing the data, as filtering decisions could disproportionately retain the flawed records, making them a non-negligible share of the analyzed subset. While the duplication issue likely stems from errors within the \texttt{DSA-TDB} itself, the incorrect moderation timestamps indicate reporting failures on the platforms’ side. These errors persist in the database despite being easily detectable with basic error checks, suggesting a lack of enforcement at the infrastructure level. Thus, researchers and policymakers must apply their own integrity checks to prevent misleading conclusions. 

\noindent\textbf{Fields usage and uninformative reporting.} The \texttt{DSA-TDB} is structured around required, conditionally required, and optional fields, providing a standardized yet flexible reporting framework. Additionally, certain fields require selecting from a predefined set of values while others allow free-text input. Among the predefined values, some are highly specific while others are broad and generic. Based on this structure, the usefulness of the database hinges on how the platforms populate these fields.

To assess the informativeness of the submitted SoRs, we analyzed the frequency with which each database attribute is used by the different platforms, distinguishing between required, conditionally required, and optional. Figure~\ref{fig:field_usage} presents our findings. As expected, all platforms consistently populate the required attributes, ensuring formal compliance with the DSA. However, the use of conditionally required attributes varies widely. While TikTok utilizes most of these attributes, the majority of platforms use almost none. Interestingly, some conditionally required attributes---such as \dbField{decision\_visibility} and \dbField{incompatible\_content\_ground}---are consistently valued across all platforms, whereas others---such as \dbField{decision\_monetary\_other}---remain entirely unused. Figure~\ref{fig:field_usage} also reveals that the optional attributes are overwhelmingly neglected. TikTok, Instagram, Facebook, and LinkedIn never populated any of them, while Pinterest, YouTube, Snapchat, and X used only a handful, in a small minority of SoRs. This issue was already discussed in early assessments of the \texttt{DSA-TDB}~\cite{trujillo2023dsa}, and our analysis confirms the lack of progress in addressing it.

One of the required attributes in the \texttt{DSA-TDB} mandates platforms to specify the type of infringement leading to a moderation action, using a set of predefined values. While some predefined categories are quite precise (e.g., pornography, harmful speech), others, such as \dbField{scope\_of\_platform\_service}, are vague catch-all labels encompassing a wide range of age, geographical, and language restrictions, disallowed goods and services, and nudity.\footnote{\url{https://transparency.dsa.ec.europa.eu/page/documentation#16-category-specification-category-category-addition-category-specification}} Early studies noted a frequent reliance on this generic category, raising concerns about the clarity of platforms' reporting~\cite{trujillo2023dsa,kaushal2024automated}. To assess whether this practice has changed, we compared the usage of the \dbField{scope\_of\_platform\_service} category in the first 100 days of the database (353M SoRs from September 25, 2023 to January 2, 2024) with the last 100 days of our observed period (573M SoRs from July 24 to October 31, 2024). Here, we refer to the former as the \textit{initial} period and to the latter as the \textit{latest} period. Figure 10 presents the results of this comparison. On average, the use of this category has remained practically unchanged, decreasing only slightly from 42.68\% to 41.04\%. However, platform-specific usage varied. Facebook, Instagram, LinkedIn, and X have increased their reliance on this generic label, while YouTube and Snapchat have significantly reduced it. TikTok and Pinterest have shown little change. These findings confirm that platforms continue to rely heavily on vague classifications, limiting the informativeness of mandatory fields. This result, combined with the rare use of optional fields, suggests that while platforms meet the formal DSA requirements, the transparency and usefulness of their reporting remain limited.

\begin{figure}[t]
    \centering
    \includegraphics[width=\columnwidth]{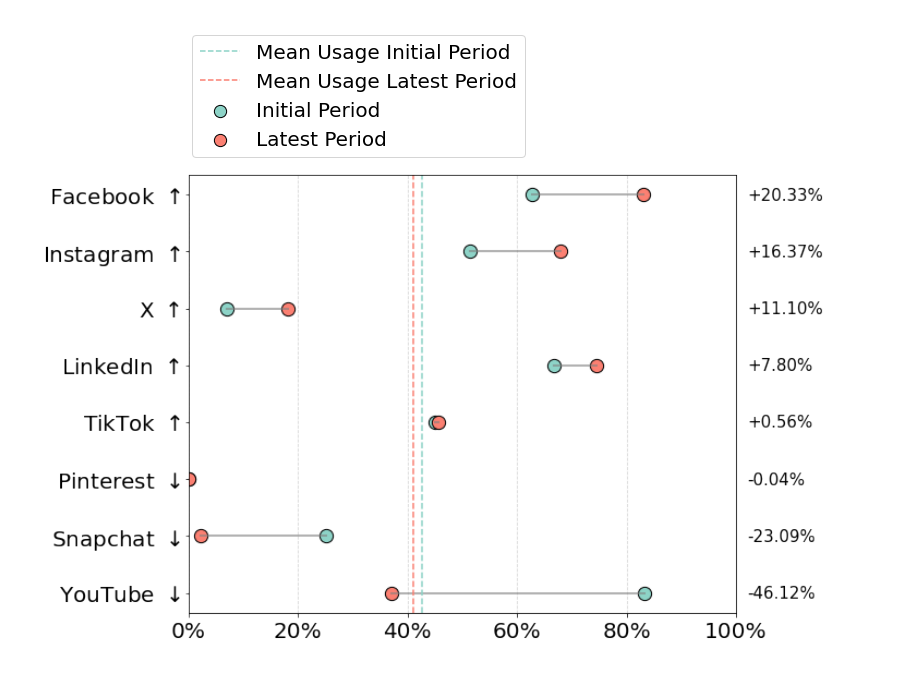}
    \caption{Frequency with which platforms resorted to the generic category \dbField{scope\_of\_platform\_service} in their SoRs. Teal dots indicate frequency of use in the initial period, while red dots indicate usage in the latest period. Horizontal bars indicate change in usage between the two periods. Platforms are shown in descending order of difference between the two periods. Arrows beside platform names indicate increases ($\uparrow$) and decreases ($\downarrow$) in frequency of use. Vertical dashed lines show mean values for each period.}
    \label{fig:scope_service}
\end{figure}

\begin{figure}[t]
    \centering
    \begin{subfigure}[b]{0.7\columnwidth}
        \centering
        \includegraphics[width=\linewidth]{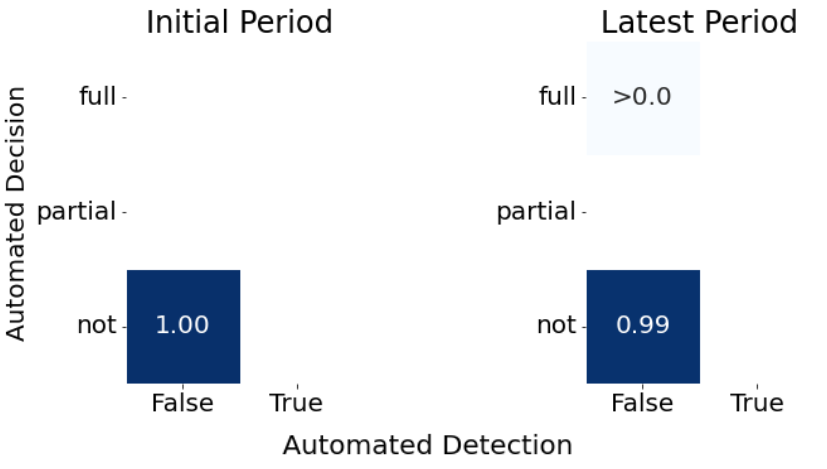}
        \caption{Use of automation.}
        \label{fig:X_automation}
    \end{subfigure}\\
    \vspace{1em}
    \begin{subfigure}[b]{0.85\columnwidth}
        \centering
        \includegraphics[width=\linewidth]{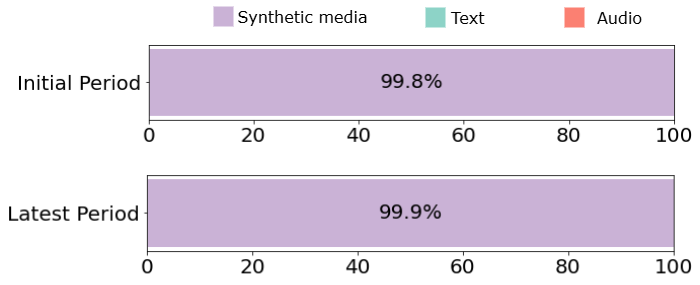}
        \caption{Type of moderated content.}
        \label{fig:X_content_type}
    \end{subfigure}
    \caption{Comparison between the use of automation (a) and the types of moderated content (b) reported by X during the initial and latest periods.}
    \label{fig:X_comparison}
\end{figure}

\noindent\textbf{Unresolved issues in X’s reporting.} In addition to identifying widespread issues with the database, previous works also highlighted some platform-specific limitations. In particular, X stood out as the platform that presented the most inconsistencies in their reporting~\cite{trujillo2023dsa,kaushal2024automated,drolsbach2024content}. Unlike all other platforms where low moderation delays are consistently linked to a high reliance on automation~\cite{papaevangelou2024content}, X continues to report near-instantaneous moderation actions while claiming to rely exclusively on manual detection and decision-making. Given that deepfakes---X’s almost exclusive target of moderation---are inherently difficult to identify manually at scale, this reporting pattern seems implausible. Further compounding the issue, past research showed that X’s \texttt{DSA-TDB} submissions contradict its own transparency reports, raising concerns about the integrity of its reports~\cite{trujillo2023dsa}. Notably, this lack of transparency was one of the factors that led the European Commission to open formal proceedings against X on December 2023~\cite{eu_proceedings_ag_x}. To assess whether X has addressed these inconsistencies, we compared its reporting from the initial and latest period. As Figures~\ref{fig:timeline} and~\ref{fig:delay_analysis} illustrate, X consistently reported zero moderation delay throughout the entire period. Additionally, Figure~\ref{fig:X_automation} shows that 99\% of X’s recent SoRs continue to indicate purely manual moderation, while Figure~\ref{fig:X_content_type} confirms its sustained focus on synthetic media. The minimal variations between the two periods are statistically non-significant ($p = 0.99$, $\chi^2$ test), indicating no meaningful changes in X’s reporting practices. These findings reaffirm the initial concerns regarding the reliability of X’s data and demonstrate a lack of improvement over time.
 \section{Discussion and Conclusions}
We analyzed 1.58B moderation actions on the the Digital Services Act Transparency Database (\texttt{DSA-TDB})---a tool designed to enhance transparency and observability in online moderation---to shed light on how major social media platforms handled content moderation during the 2024 European Parliament elections---a large-scale, multi-country political event. 
In RQ1 we sought to determine whether the \texttt{DSA-TDB} revealed changes in content moderation practices before, during, or after the electoral period. Apart from LinkedIn’s delayed enforcement against election-related misinformation, we found no evidence of meaningful shifts in moderation behavior across the eight analyzed social media platforms. This finding bears important implications. One way to explain the result is that platforms simply did not adjust their moderation strategies in response to the heightened integrity risks that are expected during political events of such magnitude~\cite{pierri2023does,majo2021role}. If true, this finding suggests that platforms either deemed their existing moderation frameworks sufficient or chose not to implement election-specific interventions. Alternatively, platforms may have indeed modified their enforcement practices, but the data available on the \texttt{DSA-TDB} was insufficient to reveal such changes. This could stem from two intertwined limitations: \textit{(i)} self-reporting deficiencies, and \textit{(ii)} structural shortcomings in the database. Since the \texttt{DSA-TDB} relies on voluntary submission, platforms may have omitted some details about their enforcement actions, either unintentionally due to internal reporting gaps, or deliberately to maintain a degree of opacity on their moderation practices. At the same time, even if platforms had fully reported their moderation actions, the database’s design may have inherently prevented the detection of moderation shifts~\cite{trujillo2023dsa}. For example, the predefined categories and mandatory fields might be too generic or too limited to capture meaningful variations in enforcement, making it difficult to track how platforms respond to evolving systemic risks like election-related disinformation.

In RQ2, we assessed the state of the \texttt{DSA-TDB} approximately one year after its launch, revisiting the initial findings that highlighted important limitations in the database and evaluating whether these issues had been mitigated or resolved. Our analysis confirms that the same shortcomings---such as incomplete reporting, vague categorization, and unreliable data, particularly from X---persist without significant improvement. This stagnation raises concerns not only about the database’s ability to serve as a robust transparency tool but also about the broader effectiveness of regulatory efforts aimed at increasing accountability in platform governance~\cite{kausche2024platform}. Transparency mechanisms like the \texttt{DSA-TDB} are only as valuable as the quality of the data they provide. If platforms systematically underuse informative fields, rely on too generic classifications, or submit records that defy plausibility, then the promise of meaningful oversight is undermined. Beyond regulatory scrutiny, these findings also speak to the role of transparency in public trust. Research has shown that users are more likely to accept and support moderation decisions when platforms provide clear, well-documented rationales~\cite{cai2024content,jhaver2019does}. Yet, the persistent opacity in platform disclosures might suggest a reluctance to fully embrace transparency, potentially reinforcing skepticism toward content moderation practices. If platforms fail to provide detailed, high-quality reports even under legal mandate, this calls into question the limits of transparency-by-design approaches and the need for stronger enforcement mechanisms to ensure compliance.

In conclusion, the implications of our study are multifold and relevant. If platforms did not adjust their moderation actions during a high-stakes election, this raises concerns about their responsiveness to systemic risks. Notably, the European Commission has already opened formal proceedings against Facebook, Instagram, and X over deficiencies in mitigating threats to civic discourse and electoral integrity, with specific reference to the 2024 European Parliament elections~\cite{eu_proceedings_ag_fb_insta,eu_proceedings_ag_x}. This regulatory scrutiny underscores the urgency of ensuring effective and transparent platform moderation practices, particularly in moments of heightened political sensitivity. Conversely, if the structure of the \texttt{DSA-TDB} prevented the surfacing of such shifts, its effectiveness as a transparency tool would be called into question. In either case, our findings suggest that the database, at least in its current form, may not yet fulfill its potential as a mechanism for scrutinizing platform behavior during politically sensitive periods.

\subsection{Limitations}
Our study relies on the quality of the \texttt{DSA-TDB} data, which is subject to the voluntary self-reporting practices of the platforms. The accuracy and completeness of the data depend on the platforms’ willingness and ability to provide detailed information, which could lead to inconsistencies or gaps in the reported actions. Additionally, our analysis was limited by the set of platforms considered in the study. While we focused on the eight major social media platforms in the EU, our findings may not fully represent the content moderation practices of other platforms. The lack of contextual information in the \texttt{DSA-TDB} presents further challenges. The database provides only metadata on moderation actions, without revealing the actual pieces of content that were moderated. As a result, it is impossible to assess the exact nature or context of the moderated content. This limitation---while necessary to protect user privacy---restricts the depth of our analyses. These factors collectively limit the ability to fully capture the effectiveness and motivations behind platform moderation actions.

\subsection{Future works}
One of the key limitations of our study was the inability to directly analyze the content that was moderated, due to the lack of content identifiers. To this end, ongoing initiatives by the European Commission aimed at designing working procedures for access to platform data under Article 40 of the DSA, could relieve the issue.\footnote{\url{https://www.eu-digital-services-act.com/Digital_Services_Act_Article_40.html}} When these become available, they could provide an opportunity to combine the self-reported records from the \texttt{DSA-TDB} with the corresponding platform data, allowing for a much richer and complete analysis of content moderation practices. Looking ahead, it would also be valuable to assess potential changes in how platforms report their moderation actions following the ongoing formal proceedings against TikTok, Facebook, Instagram, and X, which may prompt improvements in reporting practices. Furthermore, ensuring the ongoing quality of data in the \texttt{DSA-TDB} will be crucial for future transparency efforts, and our analysis could be revisited to evaluate the evolution of platform moderation during other major events, whether political or otherwise.
 
\section{Acknowledgments}
This work is supported by the ERC project DEDUCE (\textit{Data-driven and User-centered Content Moderation}) under grant \#101113826.

\bibliography{references}

\begin{table*}[t]
    \footnotesize
    \centering
    \renewcommand{\arraystretch}{1.2}
    \adjustbox{max width=\textwidth}{
    \begin{tabular}{p{0.20\textwidth}p{0.30\textwidth}p{0.5\textwidth}}
        \toprule
        \textbf{field} & \textbf{reference} & \textbf{description} \\
        \midrule
        \texttt{application date} & \S 4.1 Application Date & Indicates when a content moderation decision was applied \\
        \midrule
        \texttt{automated decision} & \S10. Automated Decision & Indicates whether the decision to moderate a content was automatic or not \\
        \midrule
        \texttt{automated detection} & \S9. Automated Detection & Indicates whether moderated content was detected automatically or not \\
        \midrule
        \texttt{category} & \S16. CATEGORY \& SPECIFICATION & Indicates the type of illegality or incompatibility with the platform's terms of services that led to a content being moderated \\
        \midrule
        \texttt{content date} & \S2.3. Date on which the content was created on the online platform & Indicates when moderated content was created \\
        \midrule
        \texttt{content type} & \S 2.1. Type of content affected & Type of the moderated content (e.g. audio, video, image, etc.) \\
        \midrule
        \texttt{content type other} & \S2.2. Specification of Content Type ``Other'' & Specification required when content type is ``other'' \\
        \midrule
        \texttt{decision ground} & \S11. Decision Grounds & Indicates whether the moderated content was deemed allegedly illegal or incompatible with the platform’s terms of service \\
        \midrule
        \texttt{incompatible content illegal} & \S12.2. Explanation of the applicability of the legal ground & Explains why a specific content has been deemed illegal according to Article 17(3)(d) of DSA \\
        \midrule
        \texttt{incompatible content ground} & \S13.1. Incompatible Content Grounds & Explains why a specific content has been deemed incompatible with the platform's terms of service \\
        \bottomrule
    \end{tabular}}
    \caption{Complete list of the DSA Transparency Database (\texttt{DSA-TDB}) fields analyzed in this study. For each field, we report its name, reference to the official documentation, and brief description.}
    \label{tab:dsa-fields}
\end{table*}
 
\section{Appendix}
\subsection{Attributes description}
Although the DSA Transparency Database provides a wide range of attributes and information, we focus our analysis on a subset of the most relevant attributes for our study. The complete list of analyzed attributes, along with references to the corresponding sections of the official documentation, and brief descriptions, is provided in Table~\ref{tab:dsa-fields}.

\subsection{Attributes values}
Platforms can assign a predefined set of values to the attributes \dbField{category}, \dbField{automated\_decision}, and \dbField{automated\_detection}. Therefore, in our analysis of moderation anomalies, we report all possible values for the automation-related attributes, as there are only two for \dbField{automated\_decision} and three for \dbField{automated\_detection}. Conversely, for the \dbField{category} attribute---which allows up to 14 values---we only include those used at least once, for the sake of brevity and clarity. The \dbField{content\_type} attribute also requires the use of a predefined set of values, including the value \dbField{other}. When a platform only used \dbField{other} as the content type, we examined the \dbField{content\_type\_other} attribute, which is a free-text attribute allowing platforms to specify content that did not fit within the predefined set of values. The specifications used in the \dbField{content\_type\_other} attribute differ from platform to platform. Consequently, we merged \dbField{content\_type} and \dbField{content\_type\_other} into a single category and reported only the values that were used at least once. The same reasoning applies to \dbField{incompatible\_content\_ground}, whose content is fully up to the platforms and not predefined.

\end{document}